\documentclass{JHEP3}

\usepackage{graphicx}
\usepackage{epsfig}
\usepackage{amsmath}
\usepackage{amssymb}
\usepackage{latexsym}

\title{Multiple photon corrections to the neutral-current
Drell-Yan process}

\author{Carlo Michel Carloni Calame$^{1,2}$, Guido Montagna$^{2,1}$,
Oreste Nicrosini$^{1,2}$ and Michele Treccani$^{2}$ \\ $^1$Istituto Nazionale 
di Fisica Nucleare, Sezione di Pavia, via A. Bassi 6, 27100, Pavia, Italy
\\ $^2$Dipartimento di Fisica
Nucleare e Teorica, Universit\`a di Pavia, via A. Bassi 6, 27100, Pavia,
Italy\\
E-mail:
\email{carlo.carloni.calame@pv.infn.it}, \email{oreste.nicrosini@pv.infn.it}, 
\email{guido.montagna@pv.infn.it}, \email{michele.treccani@pv.infn.it}}

\abstract{Precision studies of single  $W$ and $Z$ production processes at
hadron colliders require progress in the calculation of electroweak radiative
corrections. To this end, higher-order QED corrections to the neutral-current
Drell-Yan process,
due to multiple photon radiation in $Z$ leptonic decays, are calculated.
Particular attention is paid to the effects induced by such corrections on the
experimental observables which are relevant for high-precision measurements of
the $W$-boson mass at the Tevatron Run II and the LHC. The calculation is implemented
in the Monte Carlo event generator HORACE, which is available for data analysis.}

\preprint{FNT/T 2005/03}

\begin{document}

\section{Introduction}
\label{intro}
The Drell-Yan-like production of single $W$ and $Z$ bosons, with the weak boson 
decaying into a lepton pair, is a clean process with a large cross section at hadron
colliders. It is well suited for a number of precision measurements both at the
proton-antiproton ($p\bar{p}$) Fermilab Tevatron Run II and at the 
proton-proton ($pp$) CERN Large Hadron Collider (LHC) \cite{snow2001,lhc}.
While the charged-current process $pp(p\bar{p}) \to W \to l \nu_l + X$ ($l=e,\mu$)
can be used for a precise determination of the $W$-boson mass and width, the neutral-current
process $pp(p\bar{p}) \to \gamma,Z  \to l^+ l^- + X$ is of interest because $Z$ data help to
accurately calibrate detector parameters, such as the energy scale and resolution of the electromagnetic calorimeter,
which are essential ingredients for a precise determination of the $W$ mass 
\cite{snow2001,lhc,cdf,d0}. The neutral-current process can also
be used to determine the $W$ mass from the ratio of the transverse mass distributions of the
$W$ and $Z$ boson \cite{gk} (especially at high luminosity), as well as to extract the effective
weak mixing angle from the
forward-backward asymmetry \cite{cdf-afb}. Furthermore, both processes are backgrounds to many new
physics searches and can also be useful to monitor 
the collider luminosity and measure the Parton Distribution Functions \cite{lumi}. 

The success of precision studies in $W$ and $Z$ production at the Tevatron and the LHC depends
on the progress in reducing theoretical uncertainties, if possible to the 1\% level. To this end, 
QCD and electroweak corrections must be under control.  Recent developments in the theory
of $W$ and $Z$ production at hadron colliders are reviewed in Ref. \cite{pn}. 
Order $\alpha$ QED corrections to the leptonic decays of $W$ and $Z$ bosons were 
calculated in Ref. \cite{bk}. For the 
neutral-current process, which is the concern of the present study, 
the complete calculation of ${\mathcal O}(\alpha)$ QED corrections was presented in Ref.~\cite{ew-nc}, while 
the full set of one-loop electroweak corrections was computed in 
Ref. \cite{ew-nc2}.  To evaluate the impact of electroweak corrections in the experimental analysis, 
the CDF and D\O $\,$ collaborations at the Tevatron Run I made use of the 
fixed-order calculations of 
Refs. \cite{bk,bkw} for single $W$ production~\footnote{The calculation of 
Ref. \cite{bkw} refers to electroweak corrections contributing in the so-called pole approximation. 
Complete calculations of ${\mathcal O}(\alpha)$ electroweak corrections, beyond the pole
approximation, to single $W$ production are given in Ref. \cite{dk-bw}.} and of Refs. \cite{bk,ew-nc,ew-nc2} for
the single $Z$ process. However, the anticipated precision for the Drell-Yan process at the
Tevatron Run II and the LHC and, in particular, for the high-precision measurement 
of the $W$ mass, which depends on a precise determination of the $Z$ parameters, requires that leading contributions
from radiation of multiple soft and collinear photons are resummed through all orders of 
$\alpha$ \cite{pn,ew-nc}. A first attempt towards the inclusion of higher-order QED corrections 
was the calculation of $W$ and $Z$ production with radiation of two additional photons
in Ref.~\cite{bs}. Recently, higher-order corrections due to multi-photon (real and virtual)
radiation in $W$ decays at hadron colliders have been computed 
independently in Refs. \cite{pj,pv-w}. Previous investigations and Monte Carlo 
simulations of multiple QED radiation 
in $W/Z$ production in hadron collisions can be found in Ref. \cite{mpdy}. A first step towards
the inclusion of QED and QCD radiation into a single generator has been recently performed
in Ref. \cite{cy}, by combining QCD resummation with ${\mathcal O}(\alpha)$ final-state QED
corrections.

The aim of the present work is to compute higher-order
final-state QED corrections to the neutral-current Drell-Yan process 
$pp(p\bar{p}) \to \gamma,Z  \to l^+ l^- + X$ and, in particular, to quantify the
effects induced by such corrections on the experimental
observables of interest in view of a  precise determination of the $W$ mass at hadron colliders. We restrict
ourselves to consider final-state QED corrections only, because it is known from previous investigations
that electroweak corrections to Drell-Yan-like processes are largely dominated by photon radiation
from the final-state charged leptons \cite{ew-nc,ew-nc2,bkw,dk-bw}, because of the presence of large
lepton-mass logarithms arising from collinear radiation.

The present hadron collider average for the mass of the $W$ boson $M_W$ is 
$M_W = 80.454 \pm 0.059$~GeV~\cite{tevewwg}. The target $M_W$ precision is of the
order of 30 MeV (per channel per experiment) for the Tevatron Run II and of about 15 MeV for the LHC 
\cite{snow2001,lhc}, which corresponds
to a high-precision determination of $M_W$ with a relative accuracy at the $10^{-4}$ level.  
The shift in the $Z$ mass due to multi-photon radiation was estimated at the Tevatron Run I to be 
less than 10 MeV by the CDF collaboration for the $Z \to \mu^+ \mu^-$ decay~\cite{cdf} 
(when using the package PHOTOS of Ref. \cite{photos}) 
and 10 MeV by the D\O $\,$ collaboration for the $Z \to e^+ e^-$ decay~\cite{d0}. These
shifts are presently treated as systematic uncertainties on the $Z$ mass. However, in view
of the expected precision for $M_W$ in Run II and at the LHC, it will be necessary to include
multiple photon corrections into account when extracting the $W$ mass from data or when
calibrating detector components using $Z$ data. The present paper aims at completing the
recent calculations of higher-order QED corrections to single $W$ production \cite{pj,pv-w} to
cover the process of single $Z$ production, in order to eliminate (or largely reduce) 
this source of theoretical uncertainty in the experimental analysis.

The paper is organized as follows. In Sect. \ref{ta} we describe our theoretical
approach. In Sect. \ref{nr} we present the phenomenological implications of our 
study, by discussing comparisons with available calculations and quantifying the
effects of (higher-order) QED corrections to a number of distributions and observables 
of experimental interest. In Sect. \ref{concl} we give our conclusions and
perspectives.
 
\section{Theoretical approach}
\label{ta}

The observable cross section for the Drell-Yan process at hadron colliders 
with nominal squared centre of mass (c.m.) energy $s$ is obtained by convoluting 
the cross section of the parton process $q \bar{q} \to \gamma, Z \to l^+ l^-$, 
$l=e,\mu$, with the Parton Distribution Functions (PDFs) and summing over all quark flavors $q$. According to factorization theorems, it can be written as~\footnote{According to eq. (\ref{eq:ssf}), the 
transverse motion of the $Z$ boson is neglected. The modeling of $Z$ transverse momentum 
requires careful QCD calculations, including the resummation of multiple soft-gluon radiation, 
as available in the Monte Carlo program RESBOS \cite{resbos}
used at the Tevatron.}
\begin{eqnarray}
\sigma (s) \, =  && \, \sum_{q} \int d\cos\hat\theta \, dx_1\, dx_2 \, dx_3 \, dx_4 
(f_{q/A}(x_1,Q^2) \, f_{\bar{q}/B}(x_2,Q^2) + (q \leftrightarrow \bar{q}) ) \nonumber \\
&& D_{l^-} (x_3,Q^2) \, D_{l^+} (x_4,Q^2) \, \frac{d\sigma(cos\hat\theta,\hat{s})}{d\cos\hat\theta}
\label{eq:ssf}             
\end{eqnarray}
where $d\sigma(\cos\hat\theta,\hat{s})/d\cos\hat\theta$ is the parton-level differential cross section, function of
the scattering angle $\hat\theta$ in the c.m. frame and the squared parton c.m. energy 
$\hat{s} = x_1 x_2 s$. The function $f_{q(\bar{q})/(A,B)}(x_i,Q^2)$ stands for the PDFs of the initial-state
quarks(antiquarks) with momentum fractions $x_i$ ($i=1,2$; 
$0 \leq x_i \leq 1$) inside the proton(antiproton), where $(A,B)=(p,\bar{p})$ for the 
Tevatron and $(A,B)=(p,p)$ for the LHC. The quantities $D_{l^-(l^+)}(x_i,Q^2)$ represent the
QED Structure Functions (SF) for the final-state lepton(antilepton), which can be 
interpreted as the probability density of finding inside
a parent lepton a lepton with momentum fraction $x_i$ ($i=3,4$) at a virtuality scale $Q^2$ after
photon radiation \cite{sf}. The integration is over the parton momentum fractions $x_1$ and $x_2$ and
the lepton momentum fractions $x_3$ and $x_4$. This amounts to neglect the contribution of
photon radiation off the initial-state quarks, as well as the contribution of 
initial-final-state photon interference. The inclusion in eq. (\ref{eq:ssf}) of final-state photon radiation only 
is meaningful from the point of view of quantum field theory because final-state QED
corrections to a neutral-current $2 \to 2$ process form a gauge-invariant subset
within the full set of electroweak corrections and can be therefore treated separately.
Furthermore, photon radiation off (initial-state) quarks gives rise to
quark mass singularities which must be reabsorbed in PDFs, in analogy to gluon emission 
in QCD~\cite{pdf-qed}~\footnote{Very recently, the MRST group
performed a global analysis of PDFs including QED corrections and 
consistently incorporated in the parameterization of PDFs the contribution of QED radiation 
from quarks~\cite{mrst-2004}.}. After this ``renormalization" procedure,
it is known from previous investigations \cite{ew-nc,ew-nc2,bkw,dk-bw} that 
${\mathcal O}(\alpha)$ QED initial-state and initial-final-state interference
corrections to the relevant distributions are uniform and very small (at the 0.1\% level), 
while final-state photon corrections are seen to completely dominate and are responsible for
strong modifications, at the level of several \%. Consequently, initial-state and
initial-final-state interference corrections contribute very little, for instance, to $Z$-boson mass shifts,
as demonstrated in Ref. \cite{ew-nc}, and can be safely neglected
for the purposes of the present study.

To compute higher-order photon corrections and simulate the mechanism of
multi-photon emission, we make use of the
QED Parton Shower (PS) approach~\cite{ps-pv}. It consists in a numerical solution
of the QED Gribov-Lipatov-Altarelli-Parisi evolution equation 
for the lepton SF $D(x,Q^2)$ in the non-singlet channel.
The solution can be cast in the form~\cite{ps-pv}
\begin{eqnarray} 
D(x,Q^2)&=&\Pi(Q^2,m^2)\delta(1-x) \nonumber\\
&+& \bigg(\frac{\alpha}{2\pi}\bigg) \int_{m^2}^{Q^2}\Pi (Q^2,s')\frac{d s'}{s'}\Pi (s',m^2)
\int_0^{x_+} dy P(y) \delta (x-y) \nonumber\\
&+& \bigg(\frac{\alpha}{2\pi}\bigg)^2\int_{m^2}^{Q^2}\Pi (Q^2,s')
\frac{ds'}{s'}\int_{m^2}^{s'}\Pi (s',s'')\frac{ds''}{s''}
\Pi (s'',m^2)\times \nonumber \\
&&\int_0^{x_+} dx_1\int_0^{x_+}
dx_2 P(x_1)P(x_2) \delta (x-x_1x_2) + \cdots    
\label{eq:alpha2}             
\end{eqnarray}
where $P(x)$ 
\begin{equation}
P(x) \, = \, \frac{1+x^2}{1-x} - \delta(1-x) \int_0^1 \, dz \, \frac{1+z^2}{1-z}
\end{equation}
is the regularized splitting function for the QED branching $l \to l\gamma$ and 
the factor $\Pi$ 
\begin{equation}
\Pi (s_1,s_2) = \exp \left[-\frac{\alpha}{2 \pi} 
\int_{s_2}^{s_1} \frac{d s'}{s'} \int_0^{x_+} dz P(z)  \right]
\end{equation}
is the Sudakov form factor, representing the probability that
a lepton evolves from virtuality $s_2$ to virtuality $s_1$ with no emission
of photons of energy fraction greater than (an infrared regulator)
$\epsilon = 1 - x_+$. Equation (\ref{eq:alpha2}) allows to compute $D(x,Q^2)$
by means of a Monte Carlo algorithm which, as shown in detail in Ref.~\cite{ps-pv}, 
simulates the emission of a shower of (real and virtual) photons by a lepton and accounts for
exponentiation of soft photons and resummation of collinear logarithms due
to multiple hard bremsstrahlung. An ${\mathcal O}(\alpha)$ expansion of the complete PS algorithm 
can be worked out as well, as described in detail in Ref. \cite{ps-pv}, in order to compare with 
fixed-order calculations. A clear advantage of the PS algorithm 
with respect to a strictly collinear approximation is the possibility 
of generating transverse momentum $p_T$ of leptons and photons at each
branching. In the present implementation of the PS algorithm, the generation of the 
photons'  angles is performed according to the factorized part of the $Z$ radiative decay
matrix element $Z \to l^+ l^- \gamma$, i.e.
\begin{equation}
\cos\theta_\gamma\propto-\Big(\frac{p_{l^+}}{p_{l^+}\cdot k}
- \frac{p_{l^-}}{p_{l^-}\cdot k}\Big)^2
\end{equation}
where $p_{l^-(l^+)}$ is the lepton(antilepton) four-momentum and $k$ is the photon four-momentum.  
The generation of transverse degrees of freedom at each branching allows an exclusive event generation suitable to implement experimental cuts according to a realistic event selection, as shown in the next section.

\section{Numerical results and discussion}
\label{nr}

The formulation described above has been implemented in the MC event generator HORACE, developed in Ref. \cite{pv-w}
 to quantify the effect of higher-order final-state QED corrections
on the $W$-mass determination at hadron colliders. The predictions of HORACE
were recently compared with those of the independent MC generator WINHAC \cite{pj}
for various single-$W$ observables of experimental interest, finding a good agreement \cite{app}. 
However, in order to check the implementation in HORACE of QED corrections to the
neutral-current process, we performed further numerical tests in comparison with available fixed-order
calculations.  

\subsection{Comparisons with fixed-order calculations}

First, we compared the results for the ${\mathcal O}(\alpha)$ hard-bremsstrahlung correction
to the parton level cross section of Ref. \cite{bk} with the corresponding ${\mathcal O}(\alpha)$ predictions 
of HORACE. The results of this comparison are given in Tab. \ref{bak}, which shows 
the fraction of events (in \%) as a function of the lower cut-off  $k_0$ 
($k_0 = E_\gamma/E_{\rm beam}$, where 
$E_\gamma$ is the photon energy and $E_{\rm beam}$ is the parton beam energy), for both electrons and muons at a parton c.m. energy of 90 GeV.  In 
Tab. \ref{bak} the results of Berends and Kleiss calculation of Ref. \cite{bk} are
denoted as B\&K. The (absolute) differences between the predictions of the two calculations 
for the fraction of hard-photon events are 
at a few per mille level, as expected on the basis of the leading logarithmic
approximation inherent the PS algorithm. 

\begin{table}[t]
\centering
\begin{tabular}{|c|c|c|c|c|}
\multicolumn{5}{c}{}\\
\hline \raisebox{-1.5ex}[0cm][0cm]{$k_0$} &
\multicolumn{2}{|c|}{$e$} & \multicolumn{2}{|c|}{$\mu$}\\
\cline{2-5}
           & HORACE & B\&K &  HORACE & B\&K \\
\hline \hline
0.01  & 41.6 &  41.1&  22.4& 22.0\\
0.05  & 24.6 &  24.2&  13.3& 12.9\\
0.10  & 17.8 &  17.3&   9.6&  9.1\\
0.15  & 13.9 &  13.5&   7.5&  7.1\\
0.20  & 11.3 &  10.9&   6.1&  5.7\\
0.30  &  7.9 &   7.5&   4.2&  3.9\\
0.40  &  5.6 &   5.4&   3.0&  2.8\\
0.50  &  4.1 &   3.8&   2.2&  2.0\\
0.60  &  2.9 &   2.7&   1.6&  1.4\\
0.70  &  2.0 &   1.8&   1.1&  0.9\\
0.80  &  1.2 &   1.1&   0.6&  0.5\\
0.90  &  0.6 &   0.5&   0.3&  0.2\\
\hline
\end{tabular}
\caption{The fraction of events (in \%) with a photon of energy greater than 
$E_\gamma^{\rm min} = k_0 E_{\rm beam}$ as predicted
by the present MC program (HORACE) and the calculation of Ref. \cite{bk} (B\&K), at a 
parton c.m. energy of 90 GeV. 
} 
\label{bak}
\end{table}

We also performed tuned tests
 between the predictions of HORACE and those of the program ZGRAD/ZGRAD2 
  (denoted in the following as ZGRAD2) \cite{ew-nc,ew-nc2}, in 
 order to compare the results for hadron-level processes. The input parameters used in the comparisons with ZGRAD2, as well as in the
simulations described below, are the following: 
\begin{eqnarray}
&& m_e = 0.511 \times 10^{-3}~{\rm GeV} 
\qquad \, \, m_{\mu} = 0.10565836~{\rm GeV} \nonumber \\
&&\alpha^{-1} = 137.03599976  \qquad \qquad G_{\mu} = 1.16639 \times 10^{-5}~{\rm GeV}^{-2} 
\nonumber \\
&&M_Z = 91.1876~{\rm GeV} \qquad 
\sin^{2} {\theta}_{W}^{\rm \, \, eff} = 0.23150 \qquad
\Gamma_Z = 2.4952~{\rm GeV}
\end{eqnarray}
according to the latest PDG values \cite{pdg}. In terms of these parameters, we implement
in the amplitude for the parton processes $q\bar{q} \to \gamma,Z \to l^+ l^-$, with 
$q= u,d,s,c,b$, the running of the electromagnetic coupling constant, as due to fermion
loop contributions to the photon vacuum polarization~\footnote{The non-perturbative hadronic contributions to photon vacuum polarization is included in terms of the updated 
parameterization of $\Delta \alpha^{(5)}_{\rm hadrons}$ of Ref. \cite{fred-2003}.}, the running $Z$ width in
the $Z$ propagator and effective vector and axial-vector coupling constants, in agreement with
the Effective Born Approximation (EBA) described in Ref. \cite{lhc,ew-nc2}, which yields a good 
description of the complete ${\mathcal O}(\alpha)$ electroweak calculation in the region around the
$Z$ pole.  

The numerical results are obtained by imposing the following transverse momentum ($p_T$)
and pseudo-rapidity ($\eta$) cuts on the final-state leptons: 
\begin{eqnarray}
&& p_T(l^{\pm}) > 25~{\rm GeV}  \qquad \quad |\eta(l^{\pm})| < 1.2 \qquad ({\rm for \, \, the 
\, \, Tevatron}) \nonumber \\
&& p_T(l^{\pm}) > 25~{\rm GeV} \qquad \quad |\eta(l^{\pm})| < 2.4  \qquad ({\rm for \, \, the \, \, LHC})
\label{eq:cuts}
\end{eqnarray}
which model the acceptance cuts used by the experimental collaborations in the
analysis of the Drell-Yan process. In the comparison with ZGRAD2, we impose an additional
cut on the lepton-pair invariant mass $M_{l^+ l^-}$, i.e. 
$75~{\rm GeV} \leq M_{l^+ l^-} \leq 105~{\rm GeV}$, to isolate the energy 
region around the $Z$ resonance. The results refer to final-state leptons in the 
absence of lepton identification requirements, i.e. for so-called bare leptons.

\begin{table}[t]
  \medskip
  \begin{center}
  \begin{tabular}{|l||c c|c c|} \hline
     & \multicolumn{2}{c|}{Tevatron Run II} 
     & \multicolumn{2}{c|}{LHC}\\ \cline{0-4}
    Program & $e$ & $\mu$ & $e$ & $\mu$\\
    \hline
    ZGRAD2 EBA                                    & 44.11(1) & 44.11(1) &
                                                                     557.6(1) & 557.6(1) \\
    HORACE EBA                                    & 44.07(1) & 44.07(1) & 
                                                                      557.2(1) & 557.2(1)\\
    ZGRAD2                                              & 40.84(4)& 42.42(4)& 
                                                                      510(1) & 533(1)\\
    HORACE ${\mathcal O}(\alpha)$    & 40.65(3) & 42.22(1)& 
                                                                       507.2(1) & 530.4(1)\\
    HORACE   exponentiated                 & 40.60(1) & 42.21(1)& 
                                                                       507.6(1) & 530.5(1)\\
    \hline
  \end{tabular}
  \caption{Comparison between the 
present calculation (HORACE) and ZGRAD/ZGRAD2~\cite{ew-nc,ew-nc2} (denoted as 
ZGRAD2)  
for the $pp, p\bar{p} \to \gamma,Z \to l^+ l^-$, $l=e,\mu$ cross 
sections (in pb), at the c.m. energies of the Tevatron Run II ($\sqrt{s}$ = 2 TeV) 
and the LHC ($\sqrt{s}$ = 14 TeV), according to the input parameters and cuts
discussed in the text.} \label{comparison}
  \end{center}
\end{table}

The predictions for the hadron-level processes  are obtained 
by convoluting the parton-level matrix element with the
CTEQ6 PDF set~\cite{cteq}. The scale $Q^2$ is set to be
$Q^2 = \hat{s}$, $\hat{s}$ in both PDFs and lepton SFs.
The c.m. energies considered are $\sqrt{s} = 2$~TeV for the Tevatron Run II and 
$\sqrt{s} = 14$~TeV for the LHC. 

The results of the hadron-level comparison between HORACE and ZGRAD2
are shown in Tab.~\ref{comparison}, where the first two lines are the predictions 
of the programs corresponding to the cross section in the EBA, while the third and fourth lines 
refer to the QED corrected cross section in the presence of ${\mathcal O}(\alpha)$ final-state photon corrections.
For completeness, we show in the last line also the results of HORACE when including multiple QED radiation. From Tab.~\ref{comparison}, it can be seen that first-order QED corrections lower 
the cross section of about 10\% for electrons and of about 5\% for muons, both at the Tevatron and
the LHC. The effect of multiple photon radiation on the 
integrated cross section is at the 0.1\% level. The predictions of the 
two programs for the EBA cross sections are in very good agreement, being the relative differences
below the 0.1\%. Concerning the ${\mathcal O}(\alpha)$ QED corrected cross sections, the results of
the two programs differ at the 0.5\% level. As in the case of the parton-level comparison shown in 
Tab. \ref{bak}, this discrepancy can be ascribed to next-to-leading-order contributions which
are included in the full ${\mathcal O}(\alpha)$ calculation implemented in ZGRAD2 and
are missing in the PS algorithm of HORACE. 

\begin{figure}
\hskip -1.5cm
\includegraphics[height=9.5cm]{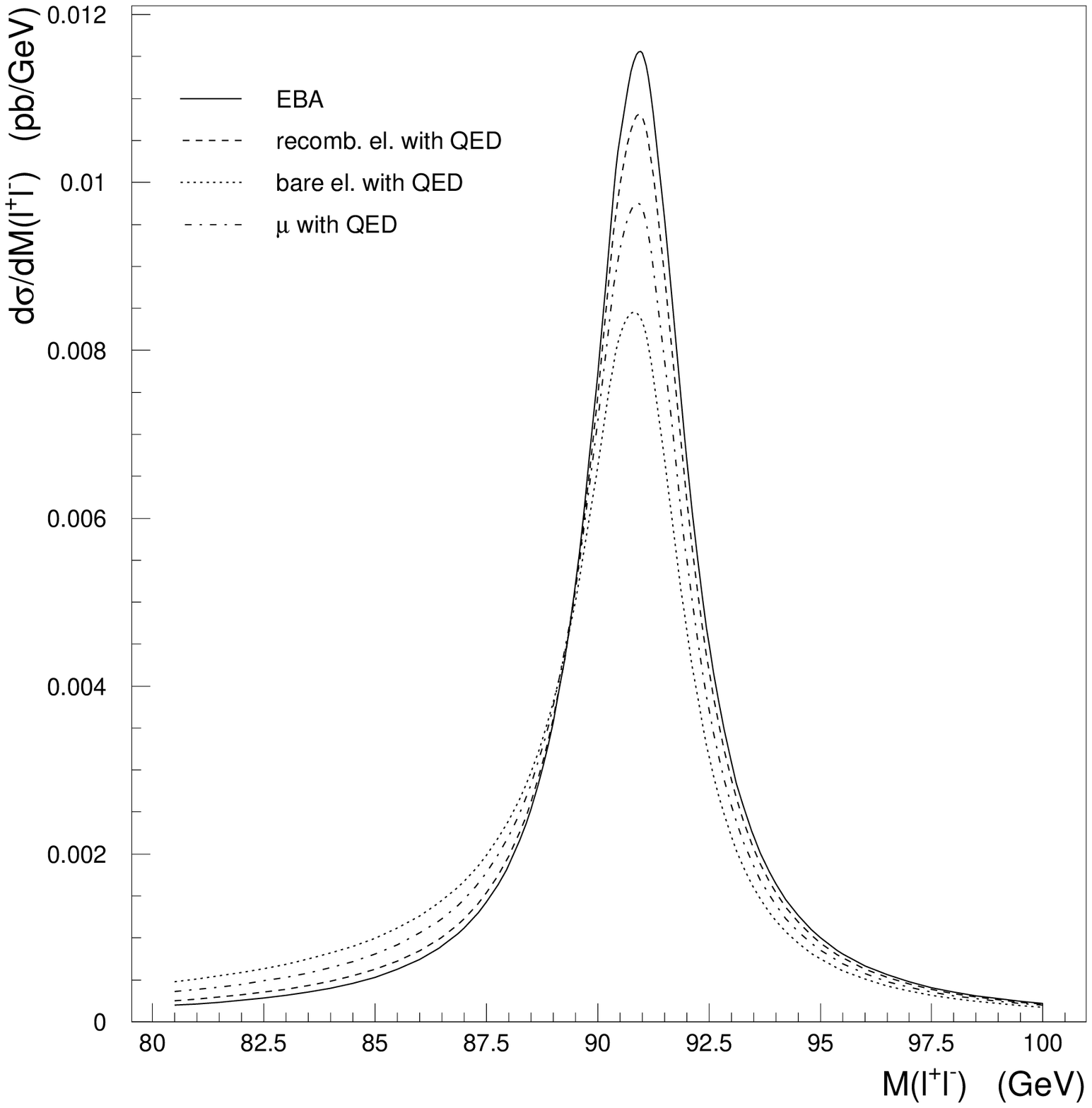}~\hskip -1.cm 
\includegraphics[height=9.5cm]{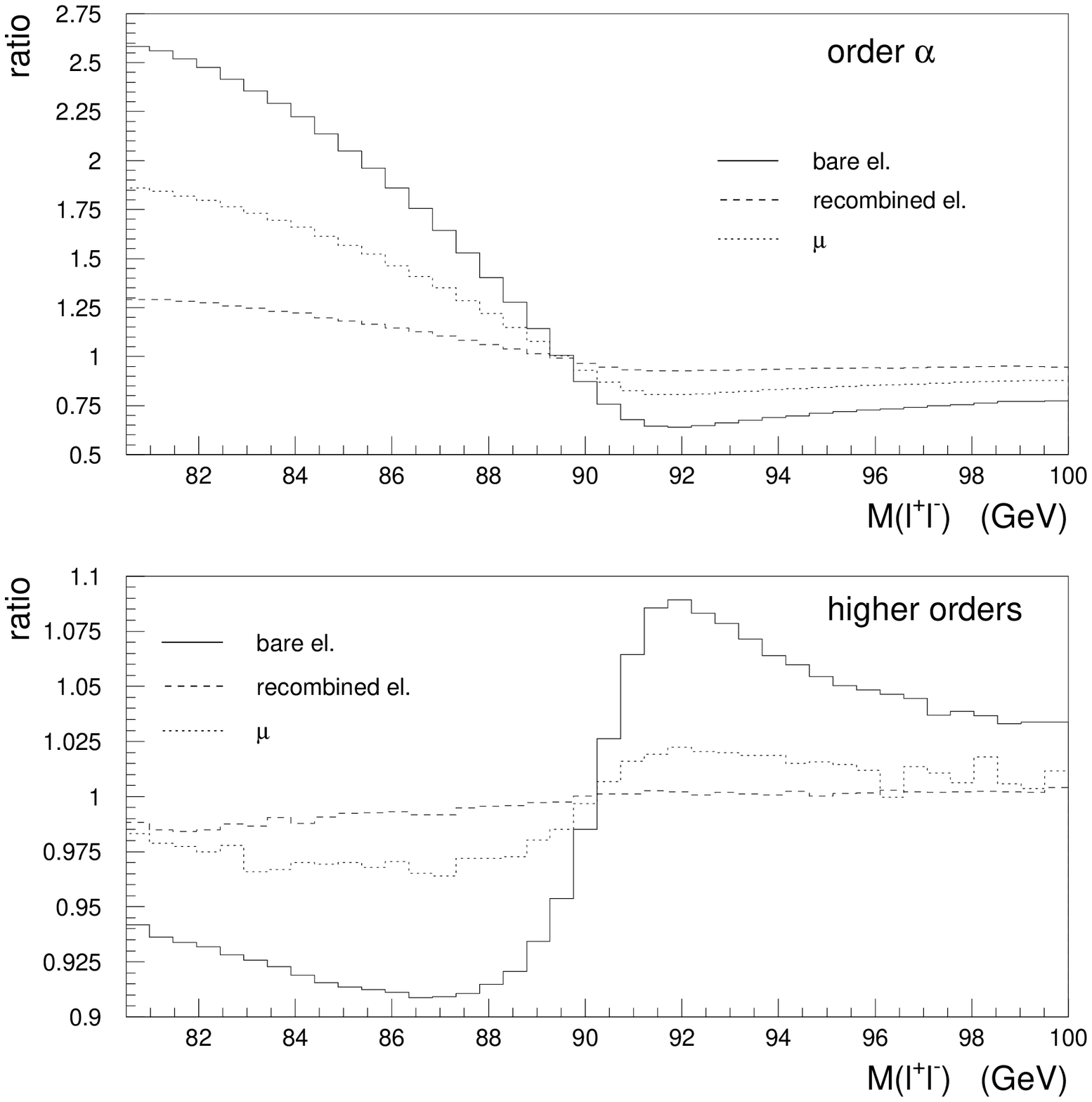}
\caption{The invariant mass distribution $M_{l^+ l^-}$ (left panel) and relative 
effect of ${\mathcal O}(\alpha)$ (right panel, up) and higher-order QED corrections 
(right panel, down) for the $Z$ 
lepton decays $Z \to e^+ e^-, \mu^+ \mu^-$ at $\sqrt{s} = 2$~TeV, according to the
lepton identification criteria discussed in the text.}
\label{invmass}
\end{figure}

\subsection{Distributions}

Having established the physical and technical accuracy of the 
implementation of the neutral-current process in HORACE, we move to the
presentation and discussion of the impact of higher-order corrections, 
with particular emphasis on the effects to the observables 
of interest for the measurement of the $W$ mass, i.e.
the lepton-pair invariant mass distribution $M_{l^+l^-} $
and the $Z$ transverse mass distribution $M_T^Z$ \cite{snow2001,lhc,cdf,d0}, which are defined as follows
\begin{eqnarray}
&& M_{l^+ l^-} = \sqrt{(E_{l^+} + E_{l^-})^2 - ({\bf p}_{l^+} + {\bf p}_{l^-})^2} \nonumber\\
&& M_T^Z = \sqrt{2 p_T(l^+) p_T(l^-) (1 - \cos\phi^{l^+l^-})}
\label{eq:obs}
\end{eqnarray}
In eq. (\ref{eq:obs}) $E_{l^-(l^+)}$ and ${\bf p}_{l^-(l^+)}$ are the lepton(antilepton) energy and three-momentum, 
$p_T(l^-),p_T(l^+)$ are the transverse momenta of the 
lepton and antilepton and $\phi^{l^+l^-}$ is the angle between the two leptons 
in the transverse plane. In order to perform a more 
realistic phenomenological
analysis and study the dependence of the QED corrections from detector
effects, we implement, in addition to
the cuts of eq. (\ref{eq:cuts}), the lepton identification requirements quoted 
in Table I of Ref.~\cite{bkw}. According to these criteria, electron and 
photon four-momenta are combined for small opening angles between
the two particles, while muons are identified as hits in the
muon chambers with an associated track consistent with a minimum 
ionizing particle. The four-momenta recombination for the electrons 
corresponds to a calorimetric particle identification, while the identification
requirements for muons are close to a bare event selection. All the numerical results
shown in the following refer to the cuts of eq. (\ref{eq:cuts}) for the Tevatron, at 
$\sqrt{s}$ = 2 TeV. 

\begin{figure}
\hskip -1.5cm \includegraphics[height=9.5cm]{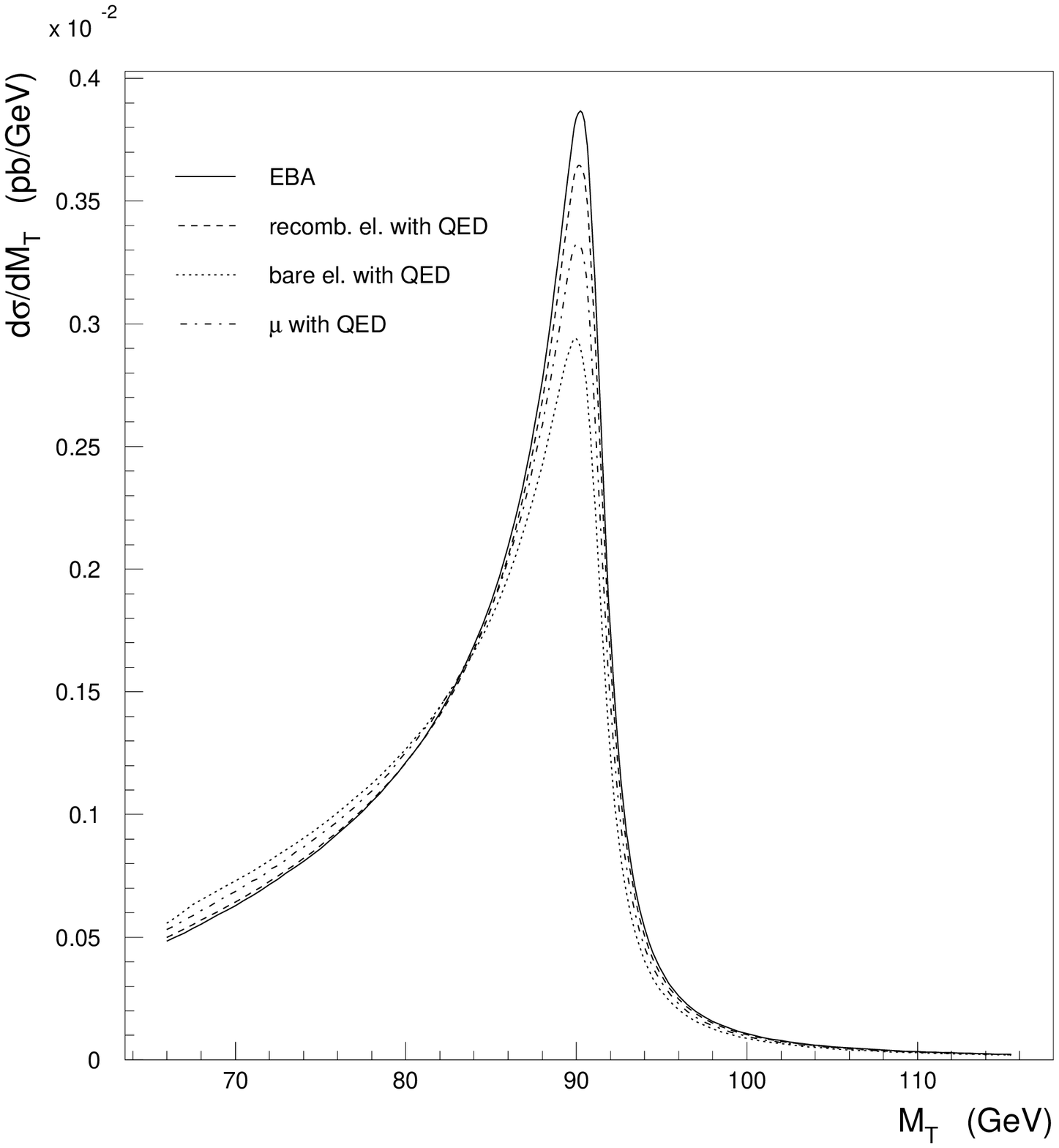}~\hskip -1.cm\includegraphics[height=9.5cm]{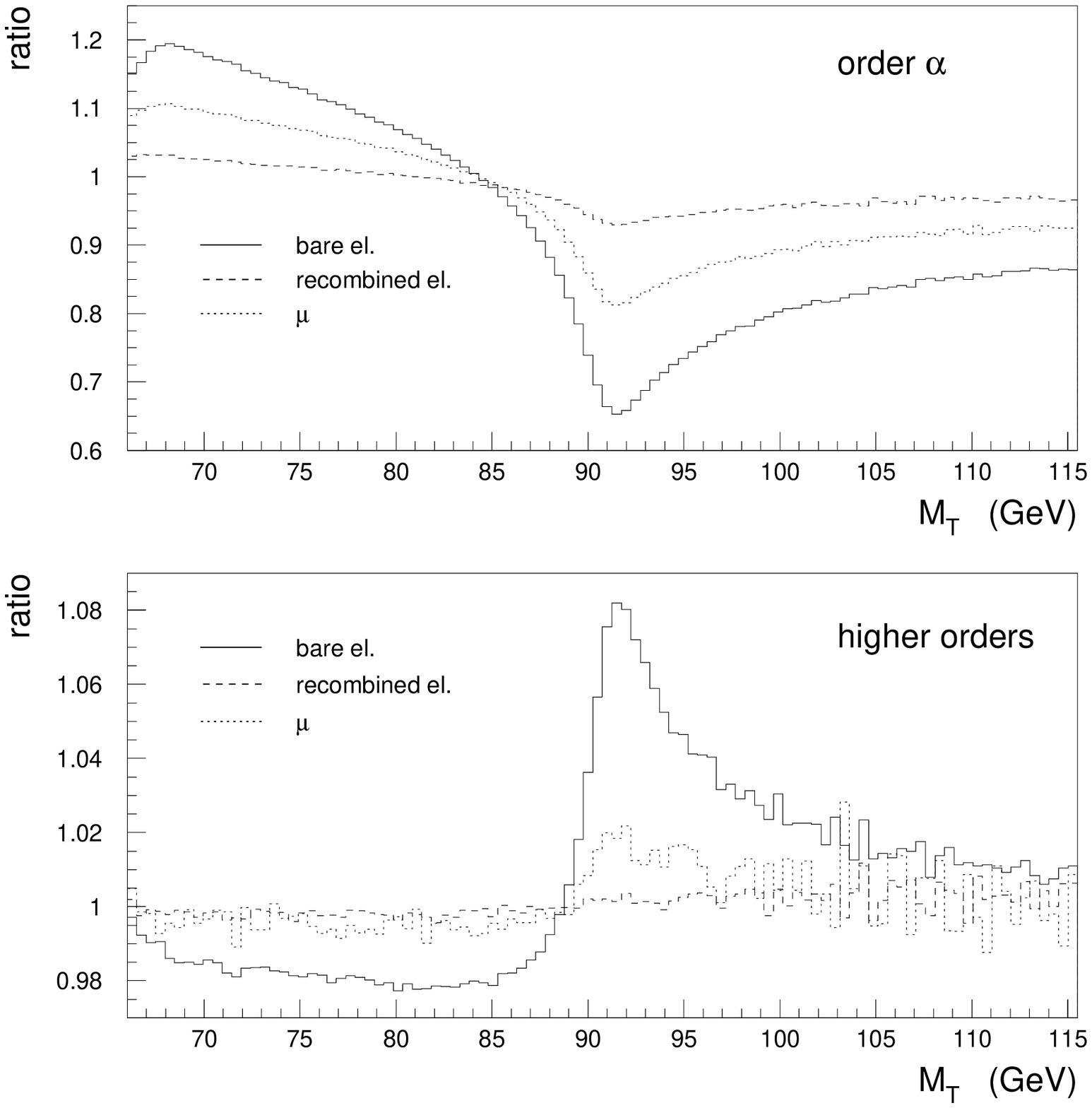}
\caption{The $Z$ transverse mass distribution $M_{T}^Z$ (left panel) and relative 
effect of ${\mathcal O}(\alpha)$ (right panel, up) and higher-order QED corrections (right panel, down) for the $Z$ 
lepton decays $Z \to e^+ e^-, \mu^+ \mu^-$ at $\sqrt{s} = 2$~TeV, according to the 
lepton identification criteria discussed in the text.}
\label{tmass}
\end{figure}

The results of our MC simulations are shown in Fig. \ref{invmass} (invariant mass) and 
Fig.~\ref{tmass} (transverse mass), when
considering for the final-state electrons and muons the lepton identification criteria discussed
above. For the sake of comparison, we also included the results corresponding to bare
electrons. The ratios shown in Fig. \ref{invmass} and in Fig. \ref{tmass} are defined as
$\sigma_{i,\alpha}/\sigma_{i,{\rm EBA}}$  and $\sigma_{i,{\rm h.o.}}/\sigma_{i,\alpha}$, 
where $\sigma_{i,{\rm EBA}}$, $\sigma_{i,\alpha}$ and $\sigma_{i,{\rm h.o.}}$ are the MC predictions for
the differential cross section at the EBA, ${\mathcal O}(\alpha)$ and higher-order level, for bin {\it i}, respectively. First of all, we notice that our results for the shape and size of ${\mathcal O}(\alpha)$ 
corrections to the invariant mass distribution are in agreement with the results of
the diagrammatic calculation of Ref. \cite{ew-nc}.  For both the invariant mass and 
transverse mass distribution and for bare leptons, the higher order corrections change sign when passing from 
below to above the $Z$ peak, and the relative correction reaches the order of 10\% for
bare electrons and of 1\% for muons, because the collinear logarithm 
$\alpha \ln(\hat{s}/m_l^2)$ is larger in the electron case. After photon recombination, the
higher-order correction for calorimetric electrons becomes flat 
and is reduced well below the 1\% level, because of the disappearance of the lepton-mass
logarithms. We also investigated the impact of higher-order corrections on the invariant mass 
distribution for large values of $M_{l^+ l^-}$, i.e. $M_{l^+ l^-} \geq 200$~GeV, which is important
in the search for new-physics and where first-order electroweak corrections amount to several 
per cent \cite{ew-nc2}. We found that higher-order QED corrections are at a few \% level and therefore
are small in comparison with other theoretical contributions, such as Sudakov-like electroweak
logarithms  \cite{ew-nc2}, and 
the expected statistical uncertainty in this invariant mass range. 

\begin{figure}
\hskip -1.5cm \includegraphics[height=9.5cm]{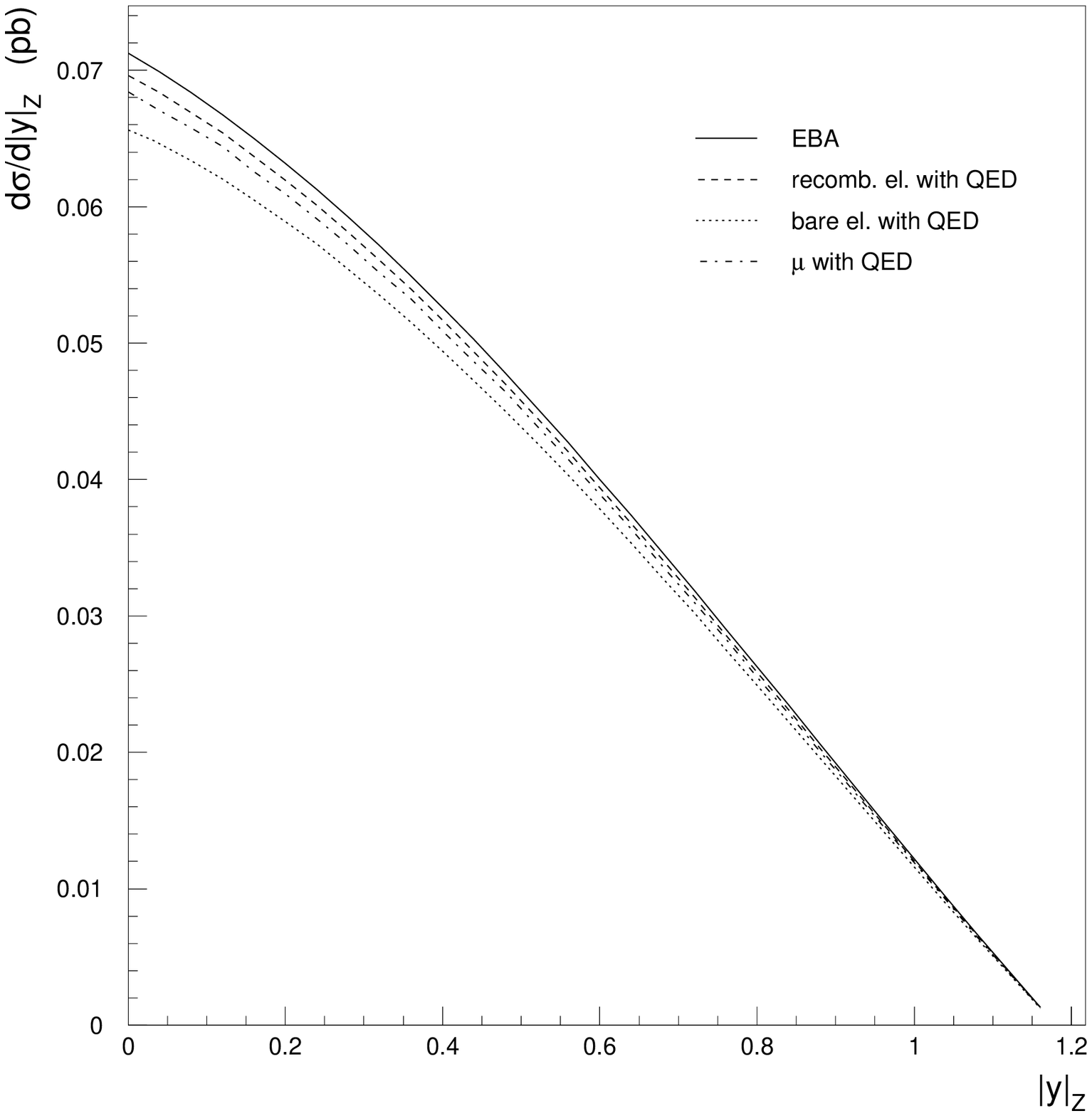}~\hskip -1.cm\includegraphics[height=9.5cm]{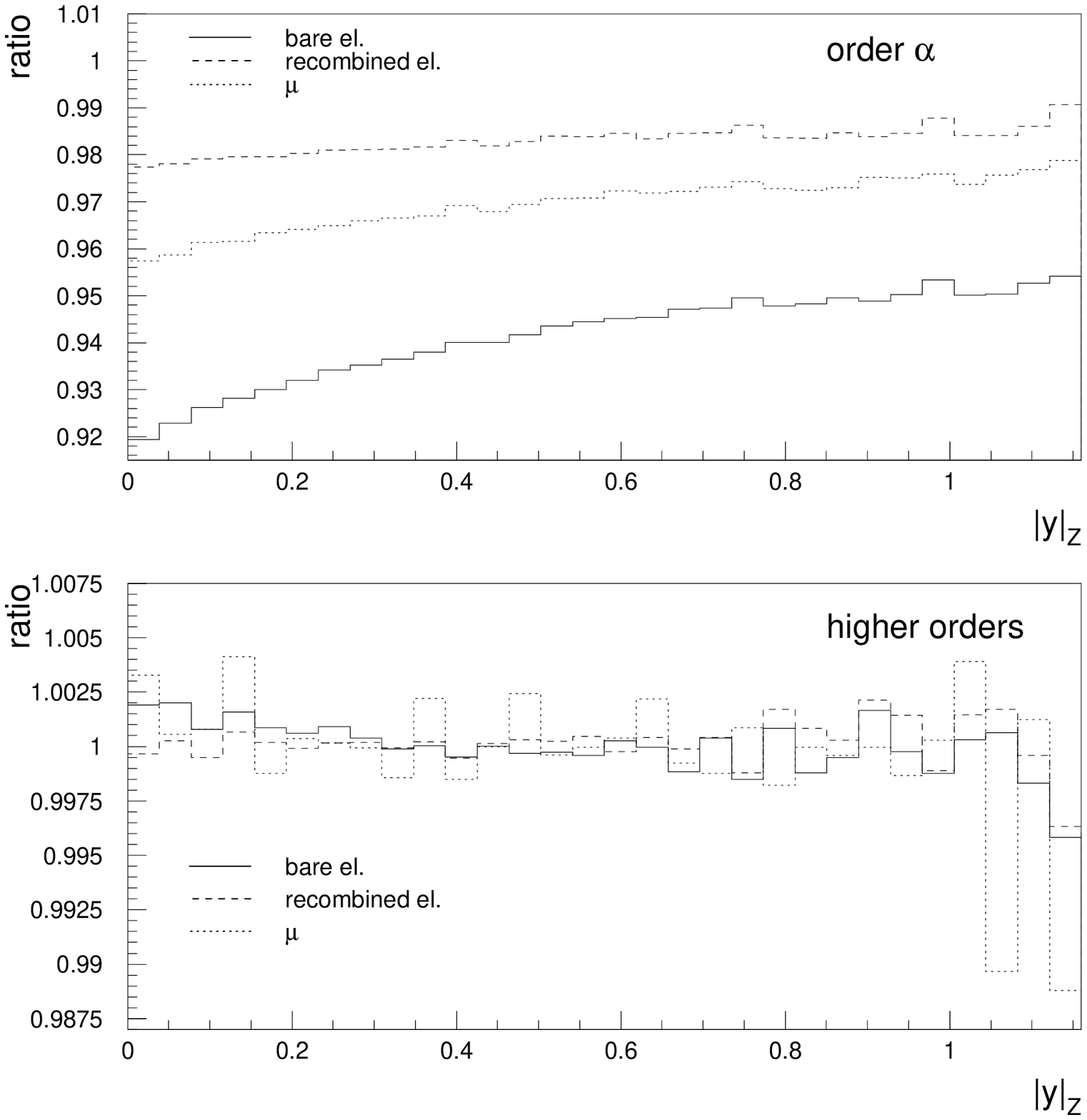}
\caption{The $Z$ rapidity distribution (left panel) and  
effect of ${\mathcal O}(\alpha)$ (right panel, up) and higher-order QED corrections (right panel, down) for the $Z$ 
lepton decays $Z \to e^+ e^-, \mu^+ \mu^-$ at $\sqrt{s} = 2$~TeV, according to the 
lepton identification criteria discussed in the text.}
\label{zp}
\end{figure}

\begin{figure}
\hskip -1.5cm \includegraphics[height=9.5cm]{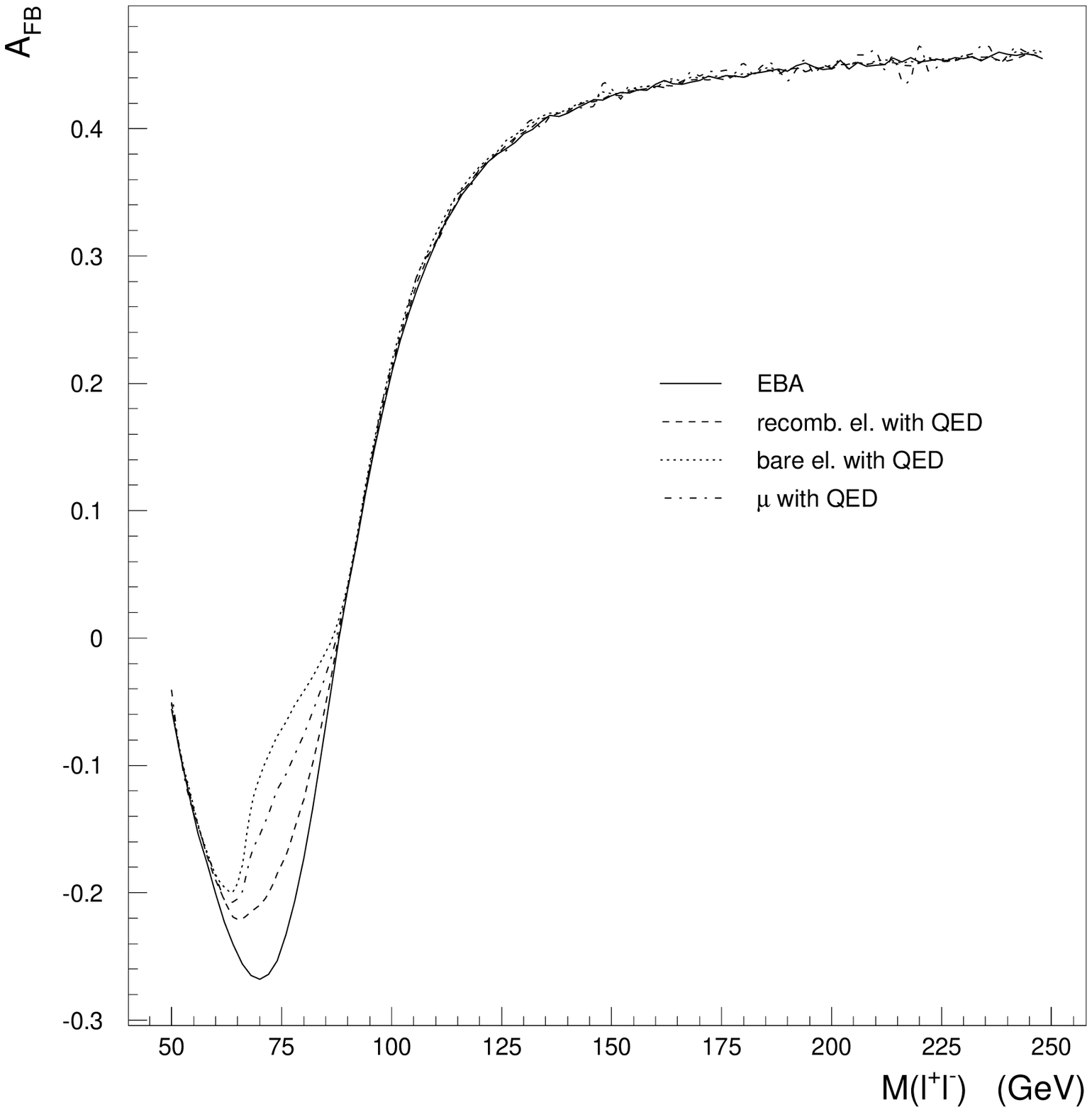}~\hskip -1.cm\includegraphics[height=9.5cm]{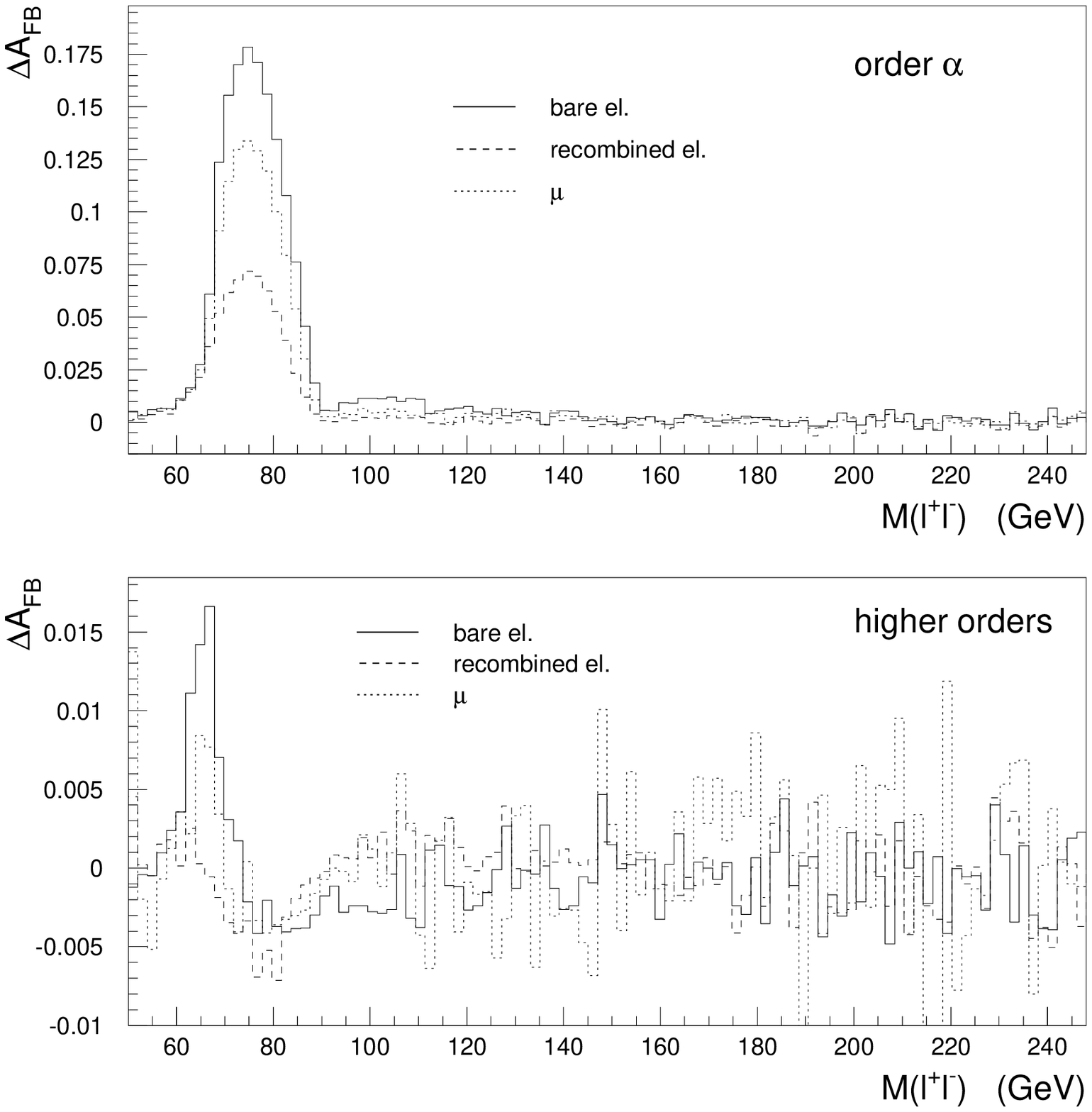}
\caption{The forward-backward asymmetry $A_{FB}$ (left panel) and  
effect of ${\mathcal O}(\alpha)$ (right panel, up) and higher-order QED corrections (right panel, down) for the $Z$ 
lepton decays $Z \to e^+ e^-, \mu^+ \mu^-$ at $\sqrt{s} = 2$~TeV, according to the 
lepton identification criteria discussed in the text.}
\label{afb}
\end{figure}

In our study, we considered other observables of interest for precision electroweak measurements, 
such as the forward-backward asymmetry \cite{lhc,cdf-afb,ew-nc,ew-nc2}, and 
the quantities relevant for luminosity monitoring, e.g. the lepton and $Z$ rapidity 
distributions \cite{lumi}. The results for the $Z$ rapidity are shown in Fig. \ref{zp}.
We observe that ${\mathcal O}(\alpha)$ QED corrections are at the level of 3-4\% for 
muons and of 1-2\% for calorimetric electrons. They are of the same order of 
next-next-to-leading order (NNLO) QCD contributions recently computed in Ref.~\cite{nnlo-qcd}.
The contribution
of higher-order corrections is quite small, at the 0.1\% level. This can 
be understood as follows. The luminosity observables are not strongly varying distributions and this
implies that ${\mathcal O}(\alpha)$ QED corrections, and consequently higher-order corrections too, are 
almost flat and smaller with respect to the case of the invariant mass and transverse mass
distributions. These results are in agreement with what observed for the analogous single-$W$
observables in Ref. \cite{app}.  The results for the forward-backward asymmetry $A_{FB}$ at the 
Tevatron Run II are shown in Fig. \ref{afb} as a function of the lepton-pair invariant mass, when 
using for $A_{FB}$ the definition for $p \bar{p}$ collisions given in Ref. 
\cite{ew-nc}. The effects of QED corrections shown in the right panel of Fig. \ref{afb} are
defined as $A_{FB}^\alpha - A_{FB}^{\rm EBA}$ and $A_{FB}^{\rm h.o.} - A_{FB}^{\alpha}$, 
giving the (absolute) contribution of ${\mathcal O}(\alpha)$ and higher-order corrections, respectively.
Our results for the ${\mathcal O}(\alpha)$ corrections well agree with those presented in 
Ref. \cite{ew-nc}, confirming that ${\mathcal O}(\alpha)$ QED corrections are large in the region below
the $Z$ peak and small around and above it. Higher-order corrections are about a factor
of ten smaller, modifying the asymmetry of about 0.01 below the peak and 0.001 above it.

\subsection{$Z$-boson mass shifts}

To complete the phenomenological analysis, we 
investigated the shift induced by multiple QED radiation on the $Z$ 
fitted mass, which is an important parameter for calibration purposes.
The strategy followed by the CDF and D\O \, collaborations to extract $M_Z$
from the data is to perform a fit to the invariant mass distribution or to the $Z$ transverse mass distribution, as a consistency check \cite{cdf,d0}. In our study, we 
perform binned $\chi^2$ fits to the $M_{l^+l^-}$ distribution, in analogy with the experimental
procedure and following the approach already used in Ref. \cite{pv-w} for the $W$ mass. 
According to this procedure, we generate 
a sample of ``pseudo-data" for the invariant mass distribution, 
calculating the distribution at the EBA level in terms of an input $Z$ mass 
$M_Z^{\rm EBA,input}$, with high numerical precision. Next, we compute 
the $M_{l^+ l^-}$ distribution including ${\mathcal O}(\alpha)$ radiative corrections for 
a number of $Z$ mass values and we
calculate, for each $M_Z$ value, the $\chi^2$ as
\begin{equation}
\chi^2 \, = \, \sum_i (\sigma_{i,\alpha} - \sigma_{i,{\rm EBA}})^2/
(\Delta\sigma_{i,\alpha}^2 + \Delta\sigma_{i,{\rm EBA}}^2)
\end{equation}
where $\sigma_{i,{\rm EBA}}$ and $\sigma_{i,\alpha}$ are the Monte Carlo
predictions at the EBA and ${\mathcal O}(\alpha)$ level, respectively, for 
bin $i$ and
$\Delta\sigma_{i,{\rm EBA}}, \Delta\sigma_{i,\alpha}$ the corresponding
statistical errors due to numerical integration. From the minimum 
of the $\chi^2$ distribution, we derive the fitted $Z$ mass value 
$M_Z^{\alpha,{\rm fitted}}$ and we quantify
the mass shift due to ${\mathcal O}(\alpha)$ corrections as $\Delta M_Z^{\alpha} \equiv 
M_Z^{\rm EBA, input} - M_Z^{\alpha,{\rm fitted}}$. The shift due to
higher-order corrections is derived according to the same procedure, 
by generating a sample of ``pseudo-data" for the $M_{l^+ l^-}$ distribution at
${\mathcal O}(\alpha)$ with an input $Z$ mass value $M_Z^{\alpha,{\rm input}}$ 
and fitting them in terms of templates of 
the invariant mass distribution obtained by including
higher-order corrections for different $M_Z$ values. The mass shift induced by higher-order 
corrections is obtained as $\Delta M_Z^{\rm h.o.} \equiv 
 M_Z^{\alpha,{\rm input}} - M_Z^{\rm h.o., fitted}$, where $M_Z^{\rm h.o., fitted}$ is the $Z$ mass value returned by the fit.

\begin{table}[t]
\centering
\begin{tabular}{|c||c|c||c||c|}
\hline
Particle&Smearing&Lepton ID&$\Delta M_Z^{\alpha}~({\rm MeV})$&$\Delta M_Z^{\rm h.o.}~({\rm MeV})$\\
\hline
$e$    &no  &no  &$-595\,(5)$&$135\,(1)$\\
$e$    &yes &no  &$-780\,(5)$&$159\,(1)$\\
$e$    &no  &yes &$-75\,(5) $&$5\,(1)  $\\
$e$    &yes &yes &$-105\,(5)$&$6\,(1)  $\\
$\mu$&no  &no  &$-270\,(5)$&$31\,(1) $\\
$\mu$&yes &no  &$-565\,(5)$&$49\,(1) $\\
$\mu$&no  &yes &$-215\,(5)$&$28\,(1) $\\
$\mu$&yes &yes &$-420\,(5)$&$44\,(1) $\\
\hline
\end{tabular}
\caption{The $Z$ mass shifts due to ${\mathcal O}(\alpha)$ ($\Delta M_Z^{\alpha}$) and higher-order
QED corrections ($\Delta M_Z^{\rm h.o.}$), at $\sqrt{s}$ = 2 TeV, according to the different experimental conditions discussed
in the text. The numbers in parentheses are statistical errors for the last digits.
} 
\label{zshift}
\end{table}

The results of
these MC experiments are given in Tab. \ref{zshift}, showing the $Z$ mass
shifts due to  ${\mathcal O}(\alpha)$ ($\Delta M_Z^{\alpha}$) and higher-order 
($\Delta M_Z^{\rm h.o.} $) corrections for
electrons and muons detected according to different combinations of experimental conditions at 
$\sqrt{s} =$~2 TeV. Indeed, in addition to the shifts obtained in the presence of the lepton identification requirements discussed above and denoted as Lepton ID in Tab. \ref{zshift}, we also 
quote the results obtained when taking into account the uncertainties in the
energy and momentum measurements of the leptons in the detector.  
These uncertainties are simulated by means of a Gaussian smearing of the particle 
four-momenta, using as standard deviation values the specifications 
relative to electrons and muons for the Run II D\O \, detector \cite{d0-det}. They are 
denoted as Smearing in Tab. \ref{zshift}. The different combinations of smearing effects 
and lepton identification criteria shown in Tab.~\ref{zshift} aim at emphasizing the
dependence of the $Z$ mass shift on detector specifications. In particular, it can be seen that
the smearing effects tend to enhance the shifts, whereas the lepton identification requirements
reduce them. When considering a specific experimental condition and comparing the 
${\mathcal O}(\alpha)$ mass shift with the higher-order one, it can be noticed that 
higher-order corrections slightly
reduce the effect due to ${\mathcal O}(\alpha)$ contributions, as expected, and that the absolute value of the
$Z$ mass shift due to multiple photon radiation is, in the most realistic case, of the order
of 10\% of that caused by one photon emission~\footnote{The opposite
sign between ${\mathcal O}(\alpha)$ and higher-order mass shifts can be understood by noticing that,
 in the vicinity of the $Z$ pole, multiple photon corrections to the invariant mass distribution 
damp the negative effect of 
${\mathcal O}(\alpha)$ corrections (see Fig.~\ref{invmass}), introducing a positive 
contribution which makes the exponentiated QED corrected distribution closer to the EBA 
distribution than the ${\mathcal O}(\alpha)$ one.}. This agrees 
with the conclusions of Ref. \cite{pv-w} for the $W$ mass shifts due to QED effects. When using the
same cuts and (simplified) detector model, we checked that the above conclusions are valid for
the LHC too, being the relative effect of QED corrections on the invariant mass distribution of the same
order as the shifts at the Tevatron. A more realistic analysis would require a full detector simulation, as well as
a modeling of the transverse motion of the $Z$, which are beyond the scope of the present paper.

\section{Conclusions and perspectives}
\label{concl}

Drell-Yan production of $Z$ bosons is an important process for precision measurements at hadron colliders. In particular, a precise extraction of the $W$ mass in a 
hadron collider environment requires a simultaneous determination of
the $Z$ mass in lepton pair production for calibration purposes. In order
to perform such precision studies at the Tevatron Run II and
the LHC, radiative corrections, including the
contribution of higher-order effects, must be under control. In this paper, we
have presented a calculation of multiple photon final-state corrections to
the neutral-current Drell-Yan process, with the aim of contributing to the
progress in reducing present theoretical uncertainties. The calculation 
is implemented in the MC program HORACE, which can be used for
data analysis. The program has been carefully tested against independent fixed-order 
calculations, finding good agreement.

We discussed the phenomenological implications of our calculation, concentrating on the
observables which are important for a precise determination of the electroweak 
parameters and for luminosity monitoring. In the presence of lepton identification 
requirements, multiple photon corrections to the lepton-pair invariant mass and to 
the $Z$ transverse mass distribution were found to modify the distributions around the $Z$ 
peak by about 0.1-1\%. For the forward-backward asymmetry, higher-order QED corrections
modify this observable by about 0.01 below the peak and 0.001 around and above it. The
quantities of interest for luminosity, such as the lepton and $Z$ rapidity distributions, are slightly affected by multiple
QED radiation, at the 0.1\% level. We also investigated the shift on the $Z$ mass, an important
parameter to calibrate detector components. The precise value of the shift was found to significantly
depend on detector effects (as expected), being the absolute value of the mass shift due to multiple radiation 
of the order of 10\% of that induced by ${\mathcal O}(\alpha)$ corrections. When associated to the 
conclusions of Refs. \cite{pj,pv-w,app} for single $W$ production, the results of the present 
paper indicate that the effects of multiple QED radiation are non-negligible in view of the
expected precision at the Tevatron Run II and the LHC and should be 
carefully considered in future 
experimental analyses. 

Concerning possible developments, it would be interesting to compare the 
predictions of HORACE with those of other programs at the level of differential
distributions and, in particular, with the photon distributions of the package PHOTOS, 
which is presently used 
at the Tevatron to estimate the impact of two-photon radiation. A second possible perspective 
would be the merging of exact ${\mathcal O}(\alpha)$ electroweak corrections with the
exclusive photon exponentiation of the PS algorithm, to improve the physical precision of 
present calculations. A further, long-term development is the combination of electroweak and QCD 
radiation into a single generator, to reach the required accuracy. 
All these developments are by now under consideration.

\acknowledgments
We are indebted to Doreen Wackeroth for useful correspondence and for her kind
collaboration in providing the ZGRAD/ZGRAD2 results quoted in the paper. We are grateful 
to U. Baur, F. Piccinini, A. Vicini and D. Wackeroth for a preliminary reading of the manuscript and 
their comments. We acknowledge useful discussions with some of the participants of the
Top/Electroweak working group of the TeV4LHC workshop.

\end{document}